\documentclass{aa}  

\usepackage{graphicx}
\usepackage{natbib}
\usepackage{url}
\bibpunct{(}{)}{;}{a}{}{,}
\usepackage{txfonts} 

\begin{document}

   \title{Revisiting the formation rate and redshift distribution of long gamma-ray bursts}

   \author{C. Kanaan
          \and
          J.A. de Freitas Pacheco}

   \institute{University of Nice-Sophia Antipolis - Observatoire de la C\^ote d'Azur\\
Laboratoire Lagrange, UMR 7293, BP 4229, F-06304 Nice cedex 4, France\\
\email{kanaan@oca.eu; pacheco@oca.eu}}

 \date{\today}

 
  \abstract{Using a novel approach, the distribution of fluences of long gamma-ray bursts derived from 
the Swift-BAT catalog was reproduced by a jet-model characterized by the distribution of the total
radiated energy in $\gamma$-rays and the distribution of the aperture angle of the emission cone. The 
best fit between simulated and observed fluence distributions permits to estimate the parameters of the model. An evolution 
of the median energy of the bursts is required to 
adequately reproduce the observed redshift distribution of the events when the formation
rate of $\gamma$-ray bursts follows the cosmic star formation rate. For our preferred 
model, the median jet energy evolves as $E_J \propto e^{0.5(1+z)}$ and the mean expected
jet energy is $3.0\times 10^{49}$ erg, which agrees with the mean value derived from
afterglow data. The estimated local formation rate is $R_{grb}=290~Gpc^{-3}yr^{-1}$, representing less 
than 9\% of the local formation rate of type Ibc supernovae. This result also suggests that the progenitors
of long gamma-ray bursts have masses $\geq 90 M_\odot$ when a Miller-Scalo initial mass
function is assumed.}
 

\maketitle

\section{Introduction}\label{intro}

Gamma-ray bursts (GRBs) are among the most violent and energetic phenomena observed in the Universe. The 
total energy emitted (if isotropically) in the form of energetic photons 
can attain values of up to few times $10^{54}$ erg, two orders of magnitude higher than the energy released by 
supernova explosions. However, this dramatic energy output can be considerably alleviated if the emission is 
strongly collimated along jets \citep{1999ApJ...525..737R}.

GRBs are classified into two main classes: long (LGRB) and short (SGRB)
\citep{1993ApJ...413L.101K}, according to the observed burst duration. The first category includes events 
with durations longer than $\sim$ 2s, while the second includes those with durations shorter than $\sim$ 2s.
More recent investigations indicate that there may be	an intermediate class of bursts that has the softest energy
spectra of the three\citep{2008A&A...489L...1H, 2010ApJ...713..552H, 2009A&A...504...67H}. The comparison between Swift and BATSE samples indicates that 
the population fraction in these classes differs because Swift detectors are more sensitive to soft and weak bursts. As a 
consequence, the probability detection of bursts included in the intermediate group is enhanced compared with the BATSE
results \citep{2010ApJ...713..552H}. Considering only the more conservative classification into two categories, Swift data
indicate that 83\% of the events are LGRBs, while only 17\% are included in the SGRB class \citep{2009A&A...504...67H}. It is
worth mentioning that these two classes have indeed some distinct properties, which are examined below.

For instance, in the plane defined by the temporal lag between 
features in the light curve observed at different wavelengths and the peak luminosity, these two 
aforementioned classes are clearly distributed separately. Moreover, there 
are other indications that LGRBs and SGRBs represent events with different origins. LGRBs 
are generally located in active star formation regions preferentially found in low-metallicity and 
low-luminosity galaxies \citep{2006Natur.441..463F}. These events might be related with type Ibc supernovae because several 
unambiguous spectroscopic identifications were obtained up to now 
\citep{2003ApJ...591L..17S, 2003Natur.423..847H, 2006ARA&A..44..507W, 2011ApJ...743..204B}.
However, late radio
observations of 68 SNIbc, searching for emission attributable to off-axis events, indicate 
that less than 10\% of type Ibc supernovae might be related to LGRBs \citep{2006ApJ...638..930S}.
Recent Swift data indicate that SGRBs occur in galaxies with different morphologies, including 
early-type objects in which no star formation activity is observed \citep{2011NewAR..55....1B, 2011MNRAS.413.2004C}.
 Moreover the positioning of these bursts indicates a wide range of spatial offsets 
from the (assumed) host galaxy \citep{2011MNRAS.413.2004C}. This would be expected if SGRBs were the consequence 
of a merger of two compact objects that have received a kick velocity during their formation 
\citep{2011MNRAS.413.2004C}. Moreover, as mentioned above, in addition to the possible existence of an
intermediate class, some authors have suggested  
two other subclasses: subluminous long GRBs (SL-LGRBs) and short GRBs with
extended emission (SGRB-EE) \citep{2005MNRAS.360L..77C, 2006ApJ...651L...5M, 2007ApJ...657L..73G, 2008MNRAS.391..405I}. 
But, possible criticisms not withstanding these subclasses are ignored in our investigation and only LGRBs are considered.
If LGRBs are the consequence of the death of massive stars, their formation rate $R_{grb}$ as a 
function of the redshift is expected to be proportional to the cosmic star formation rate (CSFR) $R_*(z)$. 
The {\it observed} rate of LGRBs, which permits estimating  the local formation rate, will depend not 
only on the CSFR, but also on the luminosity function $\phi(L)$. According to \citet{2007ApJ...656L..49S}, models 
in which LGRBs trace the CSFR and are described by a non-evolving luminosity function underestimate the 
number of events at high redshift. This difficulty could be surpassed if the luminosity function evolves 
in the sense that LGRBs are more luminous at high redshift \citep{2010ApJ...711..495B}. Including a moderate evolution 
in the luminosity function and considering an isotropic emission model, \citet{2007ApJ...656L..49S} derived a 
local LGRB formation rate $R_{grb}$=0.12 $Gpc^{-3}yr^{-1}$, while more recently, Salvaterra et al. (2011) argued 
that a strong evolution of the luminosity function is required to explain the redshift distribution 
of these events. A different method was adopted by \citet{2010MNRAS.406.1944W}, who derived the luminosity 
function and the local formation rate by directly inverting the redshift distribution of LGRBs. They 
estimated a local rate $R_{grb}$=1.3 $Gpc^{-3}yr^{-1}$, increasing up to $z\sim$ 3 and decreasing for 
high redshift. Another possibility that could explain the excess of events at high redshift 
is to assume that the efficiency of massive stars of forming GRBs increases with redshift. A variable efficiency 
of forming GRBs could be an effect due to the metallicity of the environment \citep{2006MNRAS.372.1034D, 2007ApJ...656L..49S, 2010MNRAS.406..558Q, 2011MNRAS.416.2174C, 2011MNRAS.417.3025V}.  If 
one accepts that LGRBs are formed preferentially in a metal-poor medium, the association of (at least a 
fraction of them) with type Ibc supernovae is difficult to understand since SNIbc preferentially occur in 
galaxies with high metallicity \citep{2003A&A...406..259P, 2008ApJ...673..999P}. \citet{2008A&A...491..157P}
followed a different approach to estimate the local formation rate of GRBs. They calculated the volume
detectability for sources listed in the HETE-2 catalog, that is the volume $V_{max}$ of the 
Universe in which events are bright enough to be included in their sample. Then, the 
number of GRBs inside $V_{max}$ was calculated, which permits an estimate of $R_{grb}$. They  
obtained, before correcting for beaming, a local formation rate $R_{grb}\sim$ 11 $Gpc^{-3}yr^{-1}$, a value 
substantially higher than the estimates mentioned previously. 

In the present work a novel approach was adopted to derive the local formation rate of LGRBs and the
redshift distribution of these events. A Monte Carlo code was developed to estimate the probability 
that a GRB generated at a given redshift is detected in a given energy band of the 
detector (visibility function) which in our case corresponds to the Swift-BAT experiment. 
Once the visibility function is known, the formation rate and the redshift distribution of GRBs can be 
estimated. Another particular aspect of our approach is that instead of introducing the luminosity
function of GRBs as is commonly done, we consider the energy distribution of the events, that is the 
probability for a given burst to occur with an energy $E$. The best parameters defining the assumed 
energy distribution were then determined
by comparing predicted fluences with those given by the Swift-BAT catalog. The choice of fluences is
dictated by the fact that this primary physical quantity is well defined and well measured.  
The paper is organized as follows: in Section 2 the jet 
model and the Monte Carlo code are described; in Section 3 we discuss the simulations and the data analysis 
that lead to an estimate of the parameters that define the energy distribution; in Section 4 the 
main results are given, and finally we present the conclusions in Section 5.

\section{Model and the Monte Carlo code}

Two phenomenological jet models have been intensively discussed in the literature. The so-called
universal model, in which jets have the same structure but the angular energy distribution varies
\citep{1999NewA....4..303M,2002MNRAS.332..945R} and the alternative model, which considers that the
angular energy distribution is uniform inside the jet, but the total emitted energy is nearly
the same for all events, although beamed into different opening 
angles \citep{1997ApJ...487L...1R, 2001ApJ...558L.109D,2001ApJ...562L..55F}.

In our approach, we assumed that the $\gamma$-ray emission occurs along two cones 
of aperture $2\theta$. In this case, the total energy $E_J$ emitted along the jets 
is related to the isotropic energy $E_{iso}$ by the well-known relation
\begin{equation}
\label{jet1}
E_J=E_{iso}(1-\cos\theta) ~.
\end{equation}
We assumed furthermore that the probability for an event to occur with an energy $E_J$ is given by a 
log-normal distribution, that is
\begin{equation}
\label{distributionEJ}
P(E_J)d\ln E_J = A\exp\left[-\frac{\left(\ln E_J-\ln E_{med}\right)^2}{2\sigma^2_{\ln E_j}}\right]d\ln E_J ~,
\end{equation}
where $A$ is a normalization constant. The median, $\ln E_{med}$, and the dispersion,
$\sigma_{\ln E_J}$, of the log-normal distribution are free parameters to be determined. These 
parameters are constants if there is no
evolution in the energy distribution. Since this possibility cannot be excluded, cases 
in which the median energy evolves were also considered and were modeled simply by the relation
\begin{equation}
\label{evolutionmed}
\ln E_{med}(z)=\ln E_o+\delta(1+z)~,
\end{equation}
where $\ln E_o$ and $\delta$ are constants (note that the median energy of the distribution 
at $z = 0$~ is $E_{med}(0)=E_oe^{\delta}$). 
The assumption that at high redshift LGRBs might be more energetic may be 
justified by the following argument. If jets are related to the rotation of the
GRB progenitor, it is expected that in the past, massive stars were formed preferentially in a metal-poor environment 
and, consequently, these stars have mass-loss rates lower than stars formed in metal-rich
regions. Less massive axisymmetric winds carry out less angular momentum \citep{1994Ap&SS.219..267D, 1994MNRAS.267..501D}. In this case, the progenitors of LGRBs 
formed in a metal-poor environment are expected to have higher rotation rates and to produce 
stronger jets. Note that in this context the metallicity is expected to affect the energetics
of the event, but not necessarily the formation efficiency. We cannot exclude the possibility that both the efficiency to
produce GRBs and the energy (or luminosity) distribution vary with redshift, but in this investigation we 
considered a constant efficiency per unit of stellar mass formed and possible evolutionary effects only in the energy distribution,
as modeled by eq.\ref{evolutionmed}.

We assumed also that the probability $G(\theta)$ for having a jet with an opening half-angle $\theta$ is 
independent of the jet energy. In other words, the probabilities $P(E_J)$ and $G(\theta)$ are statistically 
independent. The observed distribution of opening angles $Q(\theta)$ is related to the intrinsic distribution by the relation
\begin{equation}
\label{angledistribution}
Q(\theta)d\theta=G(\theta)\left(1-\cos\theta\right)d\theta\simeq \frac{1}{2}\theta^2G(\theta)d\theta
\end{equation}
where we used the fact that the probability for the jet with
an opening half-angle $\theta$ to be aligned with the line of sight is $(1-\cos\theta)$. From eq.~\ref{angledistribution} 
the intrinsic distribution $G(\theta)$ can be derived if the observed distribution $Q(\theta)$ is known. 

Estimates of the opening angle of the jet depend on the model parameters. \citet{2009ApJ...698...43R}
computed opening angles for 28 GRBS based on the analysis of the break time of the conical blast wave 
describing X-ray afterglows. A large sample was considered by \citet{2011arXiv1101.2458G}, who
used the Ghirlanda lower limit \citep{2004ApJ...616..331G} in the plane $\varepsilon_{peak}-fluence$,
calibrated with GRBs that have well-constrained jet opening angles. These studies suggest that
the observed distribution of values of $\theta$ can be represented by a log-normal distribution. Although this
may be only a consequence of the assumptions made for estimating $\theta$, we
adopted this in our computations, that is
\begin{equation}
\label{goldstein}
Q(\theta)d\ln\theta=\frac{1}{\sqrt{2\pi\sigma^2}}exp\left[-\frac{\left(\ln\theta-\ln\theta_{med}\right)^2}{2\sigma^2}\right]
d\ln\theta ~.
\end{equation}
Our preferred model has an observed median value for the jet opening angle $2\theta_{med}=16^o$
and a dispersion $\sigma = 0.76$. It should be mentioned that recently \citet{2012ApJ...745..168L} claimed that 
the opening angles are anti-correlated with the redshift, meaning that jets were be narrower at high $z$. This 
possibility is not considered here although it deserves more detailed investigation.

\subsection{Code}

Our code is similar to that developed by \citet{2005ApJ...620..355L}, who aimed to investigate the unified jet model. For 
the sake of completeness, we present here the main steps defining the sequence of calculations that permitted us
to produce a mock catalog of LGRBs, from which one can 
identify the differences between the present approach and that of the aforementioned authors.

\noindent

1) Initially, the redshift of the object is determined by a probability function proportional to the CSFR. The 
adopted CSFR was parametrized by the usual form proposed by \citet{2001MNRAS.326..255C}, that is
\begin{equation}
\label{csfr}
R_*(z)=h_{70}\frac{(a+bz)}{\left[1+(z/c)^d\right]} ~,
\end{equation}
where $a=0.0103$, $b=0.12$, $c=5.0$, $d=2.8$, with the CSFR given in $M_{\odot}Mpc^{-3}yr^{-1}$. $h_{70}$ is the
Hubble parameter in units of 70 $kms^{-1}Mpc^{-1}$. This expression
provides a quite good fit of the existing data (see, for instance, \citet{2006ApJ...651..142H}) and is 
consistent with the cosmological simulations by \citet{2010IJMPD..19.1233F,2011IJMPD..20.2399F}. Under these conditions, the probability $\Psi(z)$ of generating a burst in the redshift interval $z , z+dz$ is
\begin{equation}
\Psi(z)dz=N_*^{-1}\frac{R_*(z)}{(1+z)}\frac{dV}{dz}dz ~,
\end{equation}
where $dV$ is the comoving volume differential element and the normalization constant is defined by
\begin{equation}
N_*=\int^{z_{max}}_0\frac{R_*(z)}{(1+z)}\frac{dV}{dz}dz ~,
\end{equation}
where we assumed $z_{max}=15$, but the results are not significantly modified for higher values.
After obtaining the redshift $z$ from the probability function $\Psi(z)$, the luminosity distance $D_L$ is computed 
from the adopted cosmology which, in our case, was a $\Lambda$CDM model defined by the parameters 
$H_0=70~kms^{-1}Mpc^{-1}$, $\Omega_m=0.3$ and $\Omega_v=0.7$.

\noindent

2) In a second step, the energy of the jet is obtained by using the probability function $P(E_J)$ for 
given values of the median and the dispersion, according to cases including or excluding 
the evolution of the burst mean energy.

\noindent

3) The opening angle of the jet was derived from the probability function $G(\theta)$ as described previously.

\noindent

4) In the next step, we determine whether the jet points in the direction of the observer with a  probability $(1-\cos\theta)$.  
If the jet is not visible, the event is saved with its corresponding redshift as an unseen burst. If the 
event occurs in the observer's direction, we still have to verify whether the resulting 
photon rate in the observer's frame will be able to trigger the considered detector.

\noindent

5) We assumed that the photon energy distribution of the burst can be represented by a Band function 
\citep{1993ApJ...413..281B} with exponents $\alpha$ = -1.0 and $\beta$ = -2.25, typical values found from fits of 
existing data. In this case, for photon energies $\varepsilon \leq 1.25\varepsilon_{peak}$ the photon spectrum is
\begin{equation}
B(\varepsilon)=K\varepsilon^{-1}\exp\left(-\varepsilon/\varepsilon_{peak}\right)~,
\end{equation}
where $K$ is a normalization constant. For photon energies
$\varepsilon > 1.25\varepsilon_{peak}$ the photon energy distribution is
\begin{equation}
B(\varepsilon)=0.3786K\varepsilon_{peak}^{1.25}\varepsilon^{-2.25} ~,
\end{equation}
In these relations $\varepsilon_{peak}$ is the energy at which the function $\varepsilon^2B(\varepsilon)$ attains 
a maximum.
The value of $\varepsilon_{peak}$ for a given simulated event was estimated from the so-called Amati relation, which
correlates $\varepsilon_{peak}$ with the equivalent isotropic energy of the burst. In our simulations $E_{iso}$ was
estimated from eq.~1, using $E_J$ obtained in step (2).
From data by \citet{2008MNRAS.391..577A,2009A&A...508..173A} the following fit was obtained and adopted 
for our simulations (rest frame quantities):
\begin{equation}
\log\varepsilon_{peak}=-23.04+0.48\log E_{iso} \pm 0.23~,
\end{equation}
where $E_{iso}$ is in $erg$ and $\varepsilon_{peak}$ is in $keV$.

\noindent

6) According to \citet{2006ApJ...644..378B}, detectable events must have a photon flux above the threshold value $C_T$ of the detector.
\citet{2006ApJ...644..378B} computed this threshold for different experiments, in particular for Swift's BAT, whose data 
we used here. The sensitivity was computed in the $1-1000$ keV band as a 
function of the peak energy at the observer's frame ($\varepsilon_{peak}'$),
taking into account the photon distribution 
in the considered energy interval. To facilitate the computations, the sensitivity $C_T(\varepsilon_{peak}')$ was fitted by the polynomial
\begin{eqnarray}
\log C_T(\varepsilon_{peak}')=1.514-0.8801X-0.06578X^2\\ \nonumber
+0.1319X^3-0.02045X^4 ~,
\end{eqnarray}
where $X=\log\varepsilon_{peak}'$. In this equation the peak energy is given in keV and the photon flux $C_T$ is given
in $ph.cm^{-2}.s^{-1}$.

To verify wether a given simulated burst triggers the detector, its average photon flux in the $1-1000$ keV band 
was estimated by the following procedure. First, a sample including 133 LGRB with measured $T_{90}$ (duration in 
which 90\% of the total burst fluence is detected) was prepared. After correcting for the rest frame, the resulting 
distribution $\psi(T^0_{90})$ was fitted by a log-normal distribution characterized by a median
$\log T^0_{90}=1.213$ and a dispersion $\sigma_{\log T^0_{90}}=0.566$. The distribution $\psi(T^0_{90})$ permits one
to assign the duration (in the rest frame) of the considered event, via a Monte Carlo procedure. In this case, the expected 
photon rate $R_{\gamma}$ corresponding to this simulated event in the observer's frame is
\begin{equation}
\label{photonrate}
R_{\gamma}(\varepsilon_{peak})=\frac{E_JS(\varepsilon_1,\varepsilon_2)}{4\pi(1-\cos\theta)D^2_LT^0_{90}}
\end{equation}
Note that in this equation there is no (1+z) term in the denominator since it cancels with a similar term in 
the numerator due to the jet energy, which is also referred to the rest frame. The term $S(\varepsilon_1,\varepsilon_2)$
is computed from the equation
\begin{equation}
S(\varepsilon_1,\varepsilon_2)=\int^{1000(1+z)}_{(1+z)}B(\varepsilon)d\varepsilon/\int^{10^4}_1\varepsilon B
(\varepsilon)d\varepsilon ~,
\end{equation}
which represents the fraction of the total photons released in the event and detected in the range $1-1000$ keV. Note that the
limits of the integral are corrected for the rest frame and the total energy released in the event is assumed to be
restricted to the normalization interval 1~keV - 10~MeV. Under these conditions, detectable events satisfy the condition
$R_{\gamma}(\varepsilon_{peak}) \geq  C_T(\varepsilon_{peak})$ and are saved as such. It is worth mentioning that although
our detection condition refers to the average flux of the event, the BAT detector triggers the burst with respect to the peak 
flux, which is higher. In practice, this affects the estimate of the visibility function, as we see below. To 
compute peak fluxes a model for the burst time profile is required, which is currently not included in our code. The diversity 
of time profiles makes this modeling difficult, but this problem is under investigation and will be the subject of a future
paper.

\section{Simulations and data analysis}

For each pair of the parameters characterizing the log-normal distributions of the jet energy
and of the opening half-angle of the jet, that is the median and 
the dispersion, a series of runs were performed until the 
number of {\it detected} events was equal to $10^5$. The
total number of runs required to satisfy this condition depends on the adopted emission model. For 
isotropic emission, the total number is of the order on $2\times 10^6$, while in the case of beamed 
emission, the number is considerably higher, on order of $7\times 10^8$. 

The detected events constitute a catalog, including parameters 
as the redshift $z$, the jet 
energy $E_J$, the opening half-angle $\theta$ of the jet and the peak energy $\varepsilon_{peak}$, characterizing 
the hardness (or softness) of the photon spectrum. All these parameters permit one to compute the expected fluences 
in the Swift energy bands 15-25 keV, 25-50 keV, 50-100 keV, 100-150 keV, and 15-150 keV by the equation
\begin{equation}
\label{fluence}
f(\varepsilon_1,\varepsilon_2)=\frac{(1+z)E_J}{4\pi(1-\cos\theta)D^2_L}\kappa(\varepsilon_1,\varepsilon_2) ~,
\end{equation}
where the fraction of the bolometric energy in a given band ($\varepsilon_1-\varepsilon_2$) is
\begin{equation}
\kappa(\varepsilon_1,\varepsilon_2)=\int^{\varepsilon_2(1+z)}_{\varepsilon_1(1+z)}\varepsilon B(\varepsilon)d\varepsilon/
\int^{10^4}_1\varepsilon B(\varepsilon)d\varepsilon ~,
\end{equation}

After computing the fluences, they are distributed in logarithmic 
bins ($\Delta\log f(\varepsilon_i,\varepsilon_{i+1})$=0.20)
and their frequency distribution is computed. The simulated fluence frequency distribution in different energy bands
is then compared with those derived from data of the second Swift-BAT catalog \citep{2011ApJS..195....2S}. From the 475 events
present in the catalog, we prepared a culled sample of 403 objects, discarding events 
with $T_{90} <$ 2s and those without fluence data on all the considered energy bands. Of
the excluded events, 49 have short duration, 19 have incomplete data, and 4 have an uncertain classification. 
The quality of the fit was measured by the $\chi^2$ test, namely
\begin{equation}
\chi^2=\sum_i\sigma_i^{-2}\left(\nu_{i,obs}-\nu_{i,cal}\right)^2 ~,
\end{equation}
where $\nu_{i,obs}$ and $\nu_{i,cal}$ are the observed and calculated fluence frequencies in the $i$-bin respectively. 

In the next step, the code re-starts the process with a new set of parameters, which are allowed to vary within 
established intervals. The procedure is stopped when a minimum of the $\chi^2$ test is found. In reality,
the set of parameters that minimize the $\chi^2$ is not exactly the same for all energy bands. 
Assuming a Band spectrum with the same exponents $\alpha$ and $\beta$ for all simulated events certainly 
contributes to these differences and, to remedy this situation, a higher weight was given to 
the parameters determined from the widest energy band, that is 15-150 keV.

The calculations were performed by using the facilities of the computation center (SIGAMM) of the 
Observatoire de la C\^ote d'Azur. On average, about five CPU.h are necessary to optimize a pair 
of parameters for the isotropic model, while about ten CPU.h are required for the jet-model, including or 
excluding evolution of the mean energy.
  
\section{Results}  

To check our procedure, the code was tested without the beaming correction in order to compare our results
with those in the literature obtained under the same condition.
In this case, steps (3) and (4) of
the code (see Section 2.1) were skipped since there is no need to verify the alignment of the jet
with the observer's direction. 

\begin{figure}
\centering
\includegraphics[height=7cm,width=8.5cm]{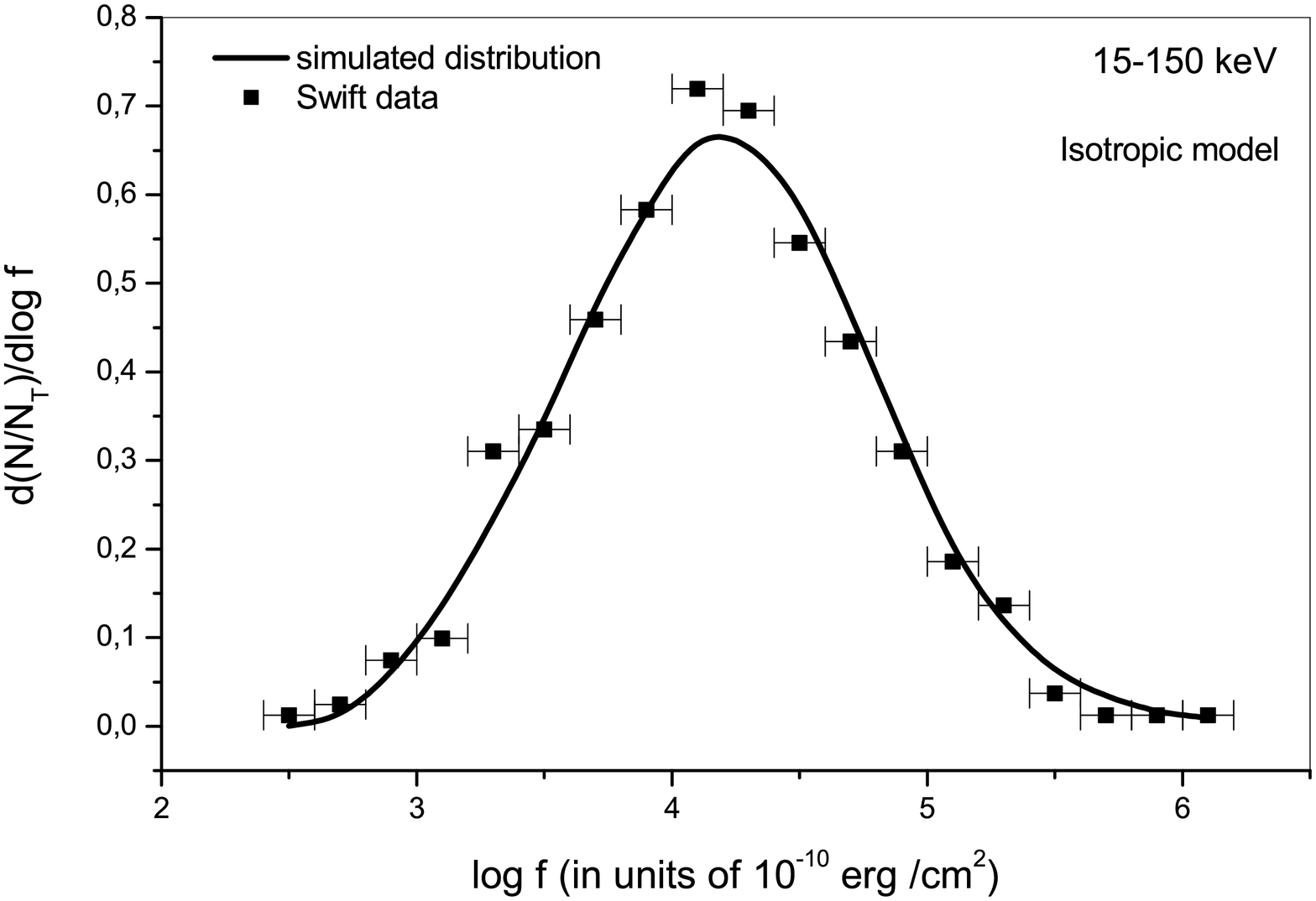}
\caption{ Frequency distribution of fluences in the energy band 15-150 keV. Points correspond to 
data derived from the Swift-BAT catalog and the error bars indicate the bin width. The solid line corresponds to simulations assuming
an isotropic emission model. The energy distribution defining the model is a log-normal distribution characterized
by a median $\log E_{iso}=50.96$ (erg) and a dispersion $\sigma_{\log E}=0.92$.}
\label{iso1}
\end{figure}
Figure 1 shows the best fit of the observed frequency distribution of fluences by simulated data in the energy 
band 15-150 keV. Simulated fluences were computed with a log-normal energy $E_{iso}$ distribution, whose median 
is $\log E_{iso} = 50.96\pm 0.34$ and whose dispersion is $\sigma_{\log E} = 0.92\pm 0.18$. Errors indicate
the uncertainties with respect to the fit of fluences in different energy bands. An early investigation
by \citet{2001ApJ...561..171J}, who have also assumed a log-normal distribution for the isotropic energy,
led to a mean burst energy of $\log E_{iso}$=53.11$\pm$0.20, about two orders of magnitude higher, but
based on a sample of only eight objects. In figure 2 the observed
frequency distribution of isotropic energies is shown in comparison with the true distribution derived
from our simulations, characterized by the parameters just mentioned. The data corresponds to 170 events listed
in the online Swift-BAT integrated parameters catalog \citep{2010ApJ...711..495B}. These data can be quite well 
fitted by a log-normal distribution with a median $\log E_{iso}=52.73$ and a dispersion $\sigma_{\log E}$=0.88. These 
results indicate that in reality, even
in the context of the isotropic emission model the mean energy released by GRB events is about 60 times 
lower than the value derived directly from observation. This difference is due to the well-known Malmquist bias, which 
favors the detection of the more 
energetic events and shifts the energy distribution toward higher values, as can be seen in figure 2. In fact,
when these selection effects are taken into account, the isotropic energy of bursts increases with the redshift
according to the analysis by \citet{2012MNRAS.423.2627W}, suggesting an evolution of the energy distribution function.

\begin{figure}
\centering
\includegraphics[height=7cm,width=8.5cm]{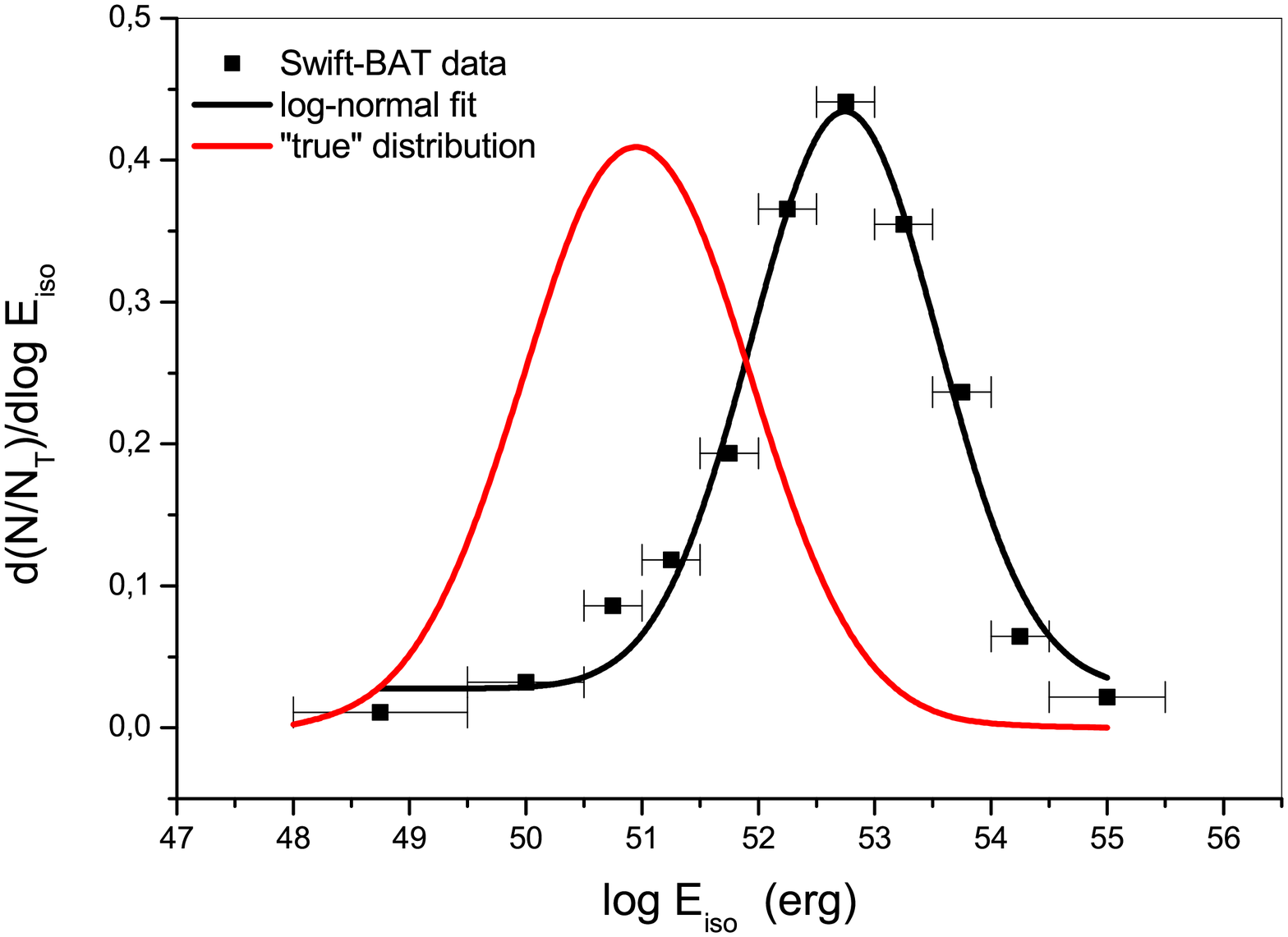}
\caption{Frequency distribution of isotropic energies. The red curve shows the true energy distribution derived
from simulations. Points represent Swift-BAT data for 170 events and error bars indicate the bin width. The solid black
curve is a log-normal fit of data.}
\label{iso2}
\end{figure}
One of the most important results derived from our simulations is the visibility function {\cal v}(z), which measures the 
fraction of bursts generated in a given redshift interval that are effectively seen by the detector, in other words, events
in the line of sight able to trigger the detector. The visibility function is essentially given by the ratio between 
bursts detected according to our rules in a given redshift interval and the total
number of bursts generated in the same redshift interval. As we emphasized above, our detection criterion
uses the average and not the peak flux. In this case, the visibility function is probably slightly underestimated.
In figure 3 the visibility function is shown for three
different sets of simulations: the isotropic model just discussed and jet models with and without energy evolution to be
analyzed below. As expected, the visibility
for the jet model without evolution is lower than that derived for the isotropic model, because not
all events are aligned with the observer. When a moderate energy evolution is 
included ($\delta$=0.5), the resulting visibility 
is slightly lower than that of the jet model without evolution up to $z\sim 1.5$ and then increases because
the events
become more energetic, being able to trigger the detector. Beyond $z \sim 7.5$ this is 
true even if a comparison is made with the isotropic model (without evolution), as can be seen in figure 3.
\begin{figure}
\centering
\includegraphics[height=7cm,width=8.5cm]{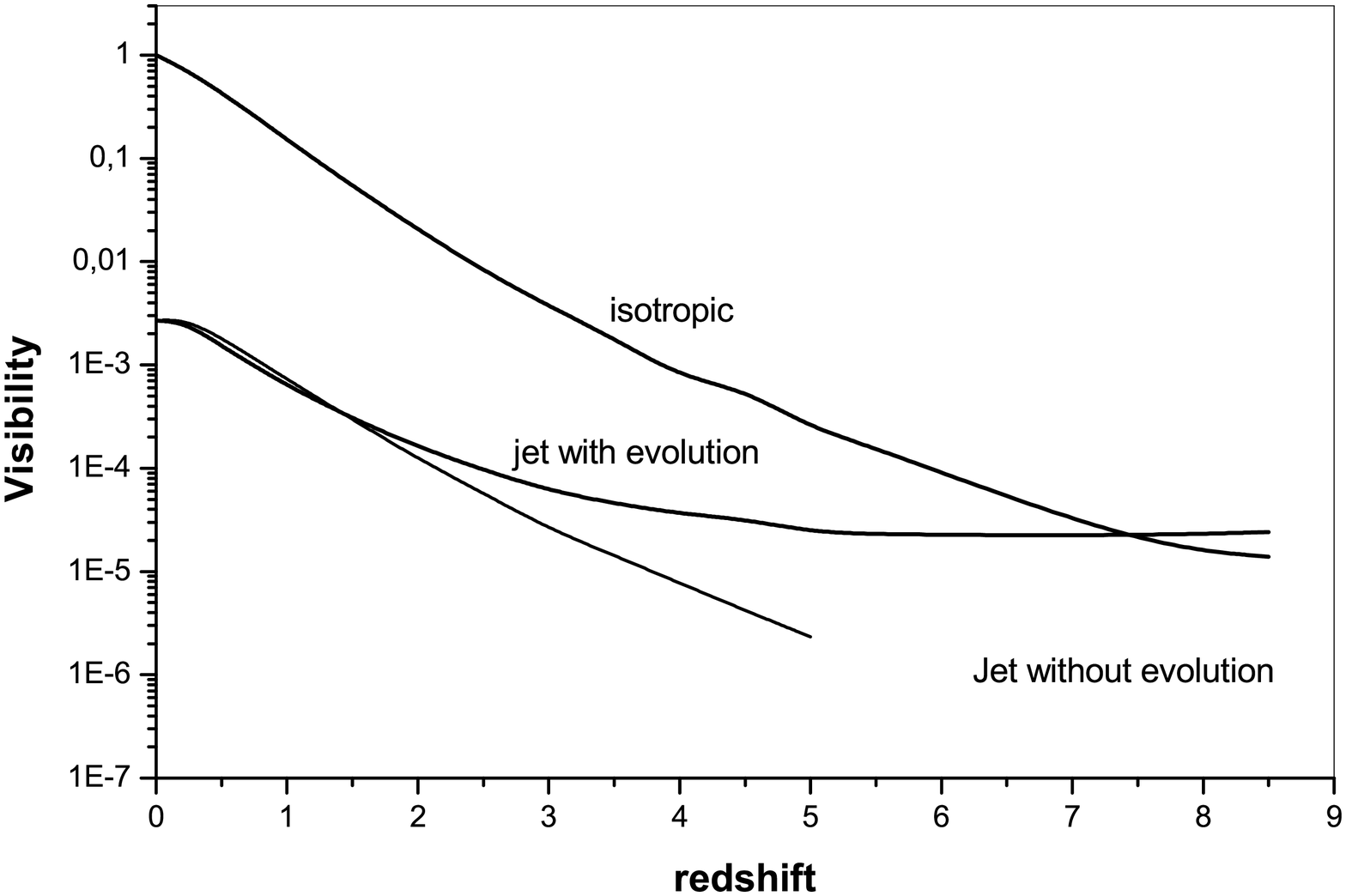}
\caption{Visibilities derived from our simulations as a function of redshift and for three
different models: isotropic emission, jet without energy evolution, and jet including evolution characterized
by a parameter $\delta$=0.5. Note that these visibilities correspond to the sensitivity of the Swift-BAT detector.}
\label{visibility}
\end{figure}
If the visibility function {\cal v}(z) is known, the expected event rate can be computed by the equation
\begin{equation}
\label{rate}
\frac{dT_{grb}}{dt} = K_{grb}\int_0^{z_{max}}\frac{R_*(z)}{(1+z)}\frac{v(z)r^2(z)dz}
{\sqrt{\Omega_v+\Omega_m(1+z)^3}} ~,
\end{equation}
where $K_{grb}=4\pi\lambda_{grb}(c/H_0)^3=9.89\times 10^{11}\lambda_{grb}$. The numerical value is obtained
if the fraction of mass of the formed stars originating in GRB events ($\lambda_{grb}$) is given in $M_{\odot}^{-1}$, the 
CSFR ($R_*(z)$) is given in $M_{\odot}Mpc^{-3}yr^{-1}$, and the event rate
$dT_{grb}/dt$ is in $yr^{-1}$. Moreover, in the equation above, $r(z)$ is the comoving distance in units of the Hubble radius
given by
\begin{equation}
r(z)=\int_0^z\frac{dx}{\sqrt{\Omega_v+\Omega_m(1+x)^3}} ~.
\end{equation}
For the considered energy distribution, the numerical value of the
integral in eq.~\ref{rate} is $4.71\times 10^{-3}$, implying a GRB rate of 
$dT_{grb}/dt=4.66\times 10^9\lambda_{grb}~yr^{-1}$.
Swift data indicate an all-sky frequency of 2.45 bursts/day \citep{2004ApJ...611.1005G} and, using the observed fraction of
long GRBs, the observed frequency of these events is 2.0 bursts/day. Comparing this with our theoretical rate, this implies
$\lambda_{grb}=1.57\times 10^{-7}~M_{\odot}^{-1}$. This value should be compared with the value derived by \citet{2007ApJ...656L..49S}
$\lambda_{grb}=1.14\times 10^{-8}~M_{\odot}^{-1}$ based on similar assumptions, that is isotropic emission and no evolution
on the burst energetics. Using our adopted local star formation rate of 
0.0103 $M_{\odot}Mpc^{-3}yr^{-1}$, one obtains a local formation rate for LGRBs of $R_{grb}$=1.6 $Gpc^{-3}yr^{-1}$.
Despite the common assumptions (isotropy and no evolution),
\citet{2007ApJ...656L..49S} adopted a different procedure to compute the local formation rate: their analysis
was based on the burst luminosity function, while we considered the bolometric energy distribution. Moreover, our burst detection 
criterion is quite different, which may explain the difference of one order
of magnitude between the two estimates of the local formation rate.

\subsection{Jet models: local formation rates}

Jet models were investigated under two main assumptions. In the first series of 
runs, the median of the jet energy distribution was kept constant, while in the second series of models 
the median was allowed to evolve with redshift according to eq.~\ref{evolutionmed}. In this case, for a given 
value of the parameter $\delta$, the code searches for the best fit of the observed fluence distribution by varying the median and the dispersion of
the log-normal representing the true jet energy distribution, following the same procedure as described above.
\begin{figure}
\centering
\includegraphics[height=7cm,width=8.5cm]{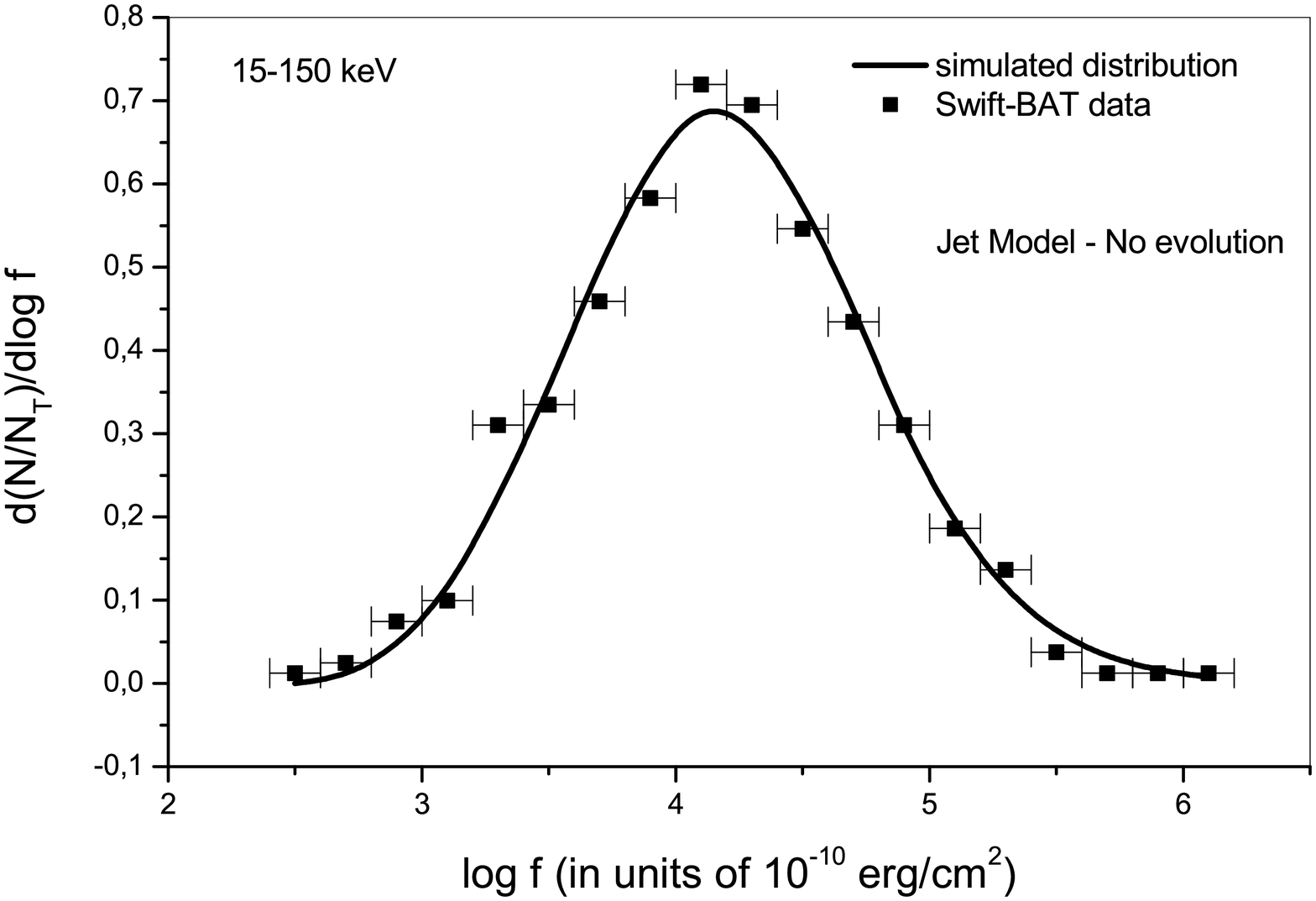}
\caption{Fluence frequency distribution for the jet model without energy evolution. Points correspond to Swift-BAT
data as in fig.1. The solid curve was derived from simulations using a log-normal for the jet energy distribution defined
by a median $\log E_J=49.26$ (erg) and a dispersion $\sigma_{\log E}=0.43$.}
\label{je1}
\end{figure}
Figure 4 shows the best fit of the observed fluence frequency distribution by simulated data based on the jet model
without evolution in the energy distribution. In this case, the model providing the best fit corresponds to a log-normal
jet energy distribution with a median $\log E_J= 49.26\pm 0.18$ and a dispersion $\sigma_{\log E_J}=0.43\pm 0.21$. Note that
the distribution energy of the jet model is narrower than that derived for the isotropic model. As expected, the jet model 
implies an average energy released in the form of $\gamma$-rays about 54 times lower than the isotropic model. The 
reduction factor is still higher (about 3200) when the comparison is made with the mean isotropic value deduced directly 
from observations, alleviating, as expected, the energy budget of these explosive events.

For a jet model without evolution the numerical integral in eq.~\ref{rate} is $2.35\times 10^{-5}$, leading to
$\lambda_{grb}=3.14\times 10^{-5}~M_{\odot}^{-1}$ and a local GRB formation rate of 324 $Gpc^{-3}yr^{-1}$, which is about 200
times higher than the rate derived under the assumption that the emission is isotropic.

When evolutionary effects are included in the energy distribution, the best fit of the observed frequency 
distribution of fluences for $\delta$=0.5 was obtained with a log-normal distribution defined by the parameters 
$\log E_0=48.46\pm 0.26$ and $\sigma_{\log E_J}=0.39\pm0.24$, as shown in figure 5. In this 
particular model, the mean jet energy at 
$z$=1 is $\log E_J$=49.18, while at $z$=7 the mean value is $\log E_J$=50.48, that is about 20 times higher. In figure 6 the 
frequency distribution of jet energies corresponding to the detected events is shown. Note that the distribution
is not symmetric, but has an extended high-energy wing because of the evolution of the 
energy distribution.
From this distribution the mean expected value of the jet energy is $\log E_J$=49.48. It is worth mentioning 
that this value derived from
simulations compares quite well with the mean jet energy resulting from the sample of GRBs with prominent jet
breaks by \citet{2009ApJ...698...43R}, that is $\log E_J$=49.64 and with the mean jet energy derived
by \citet{2001ApJ...552...36S}, that is $\log E_J$=49.69.
The numerical integral in eq.~\ref{rate} is in this particular case equal to 
$2.62\times 10^{-5}$, resulting in a formation fraction of GRBs of $\lambda_{grb}=2.82\times 10^{-5}~M_{\odot}^{-1}$
and in a local formation rate of 290 $Gpc^{-3}yr^{-3}$. Thus, the introduction of a moderate evolution in the jet energy
distribution does not significantly change the estimate of the local formation rate. It is interesting  
that Lu at al. (2012) also found a similar local formation rate
($R_{grb}$=285 $Gpc^{-3}yr^{-1}$), but their analysis is based on a possible anti-correlation between 
the opening angle and the redshift.
\begin{figure}
\centering
\includegraphics[height=7cm,width=8.5cm]{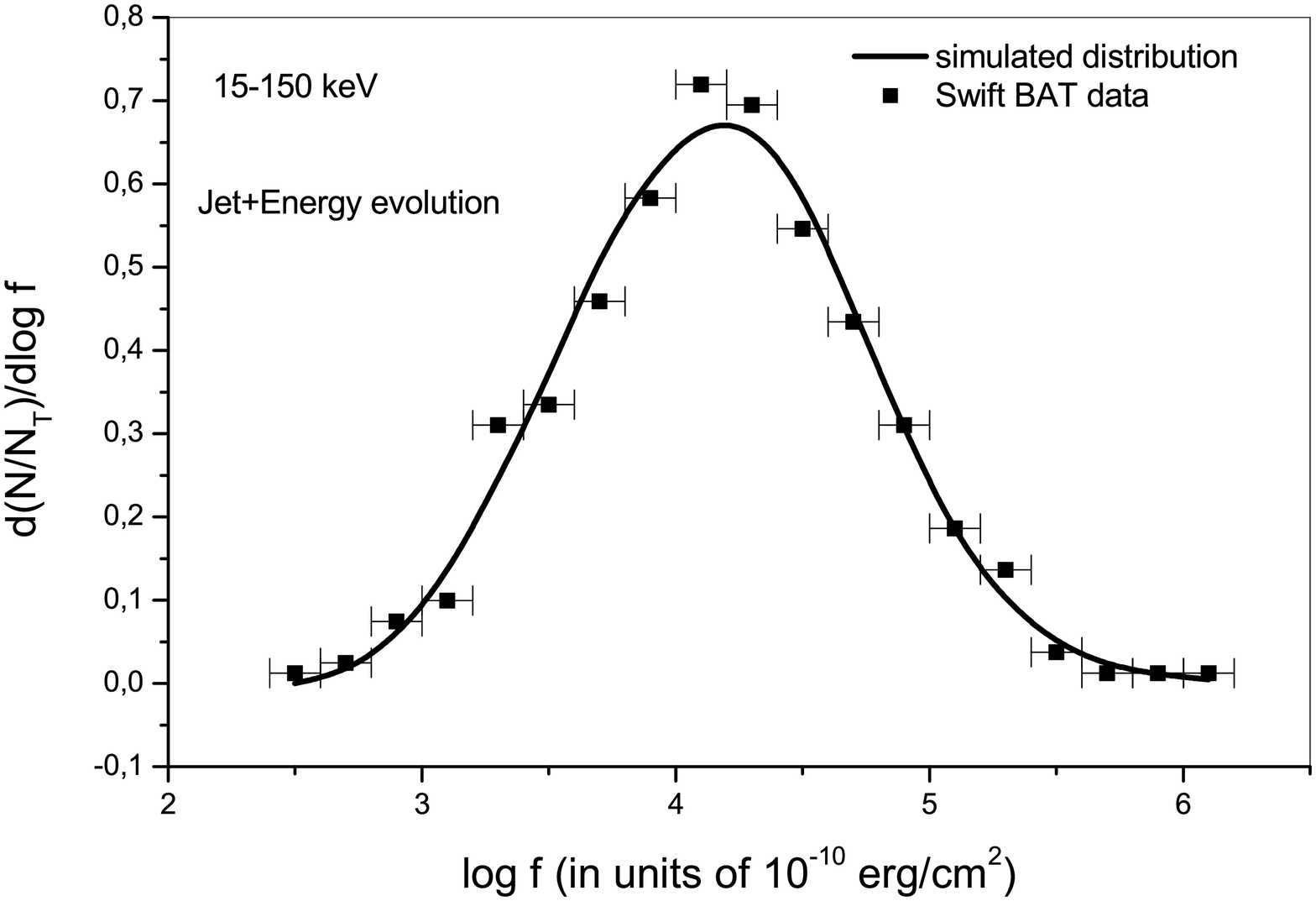}
\caption{Fluence frequency distribution for the jet model including energy evolution ($\delta$=0.5). Points 
correspond to Swift-BAT data as in fig.1. The solid curve was derived from simulations using a log-normal distribution
for the jet energy distribution in which the median varies linearly with the redshift.}
\label{jt2}
\end{figure}

\begin{figure}
\centering
\includegraphics[height=7cm,width=8.5cm]{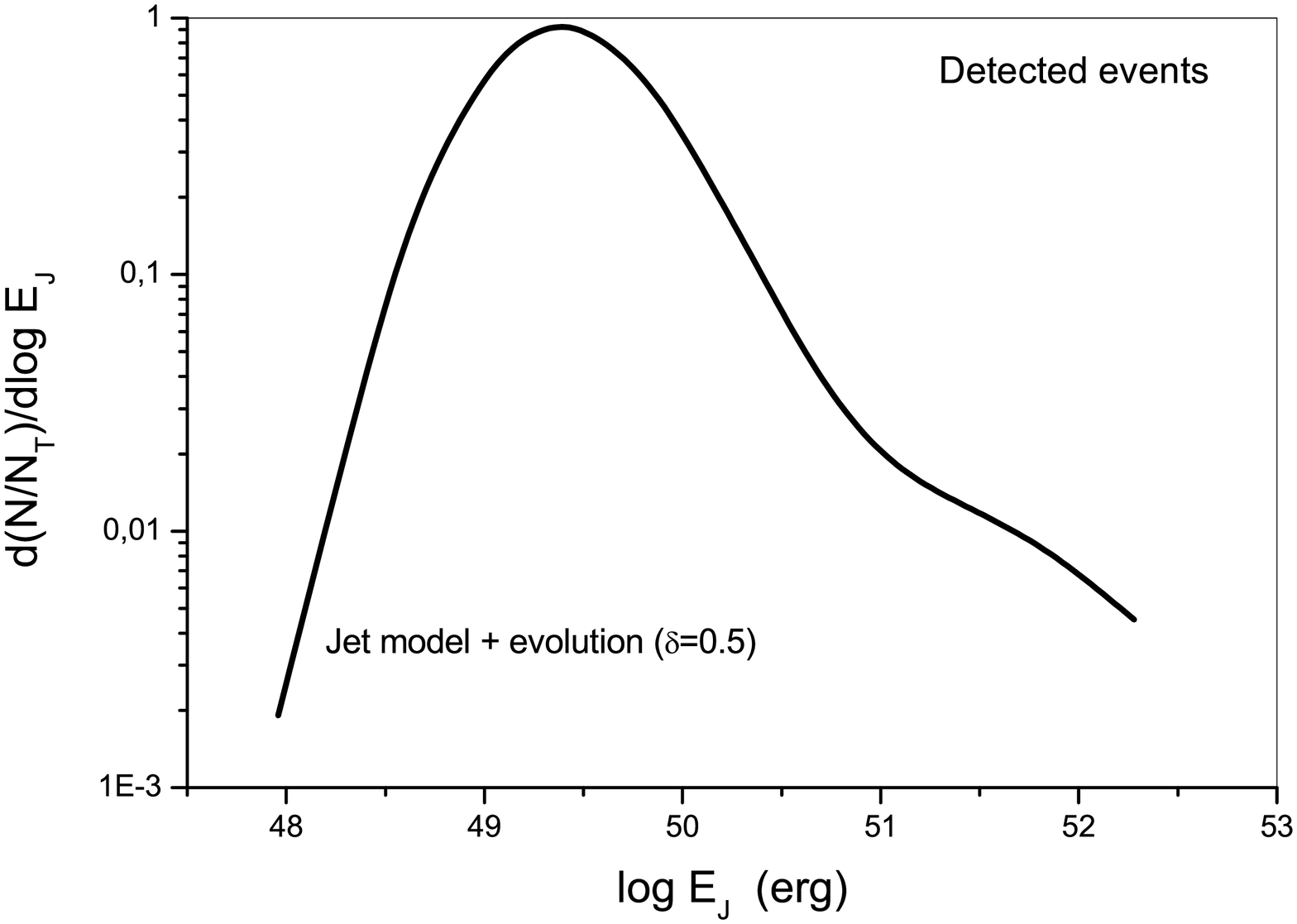}
\caption{Frequency distribution of energies for the jet model including evolution ($\delta$=0.5). The
distribution corresponds to detected events only.}
\label{jet3}
\end{figure}

\subsection{Redshift distribution}

Previous investigations on the redshift distribution of GRBs led to the conclusion that without
evolution either in the luminosity function or/and in the formation efficiency of GRBs,
the number of events at high redshift is underestimated if the formation rate simply follows
the CSFR. Since then, the number of GRBs with measured redshift has increased considerably,
permitting more robust analyses. To compare the present data with our simulations,
the redshift frequency distribution was calculated using 189 events given in the online Swift-BAT 
catalog of integrated parameters \citep{Butler:2010:Misc}. 

Considering the difficulty to model the selection effects that affect the detection of the GRB host galaxy and
the redshift measurement, they are not considered here. In this case, defining $N_{grb}=dT_{grb}/dt$, the 
expected redshift distribution resulting from our simulations is
\begin{equation}
\label{zdistribution}
\frac{d\nu_{grb}}{dz}=\frac{1}{N_{grb}}\frac{dN_{grb}(z)}{dz} ~.
\end{equation}

\begin{figure}
\centering
\includegraphics[height=7cm,width=8.5cm]{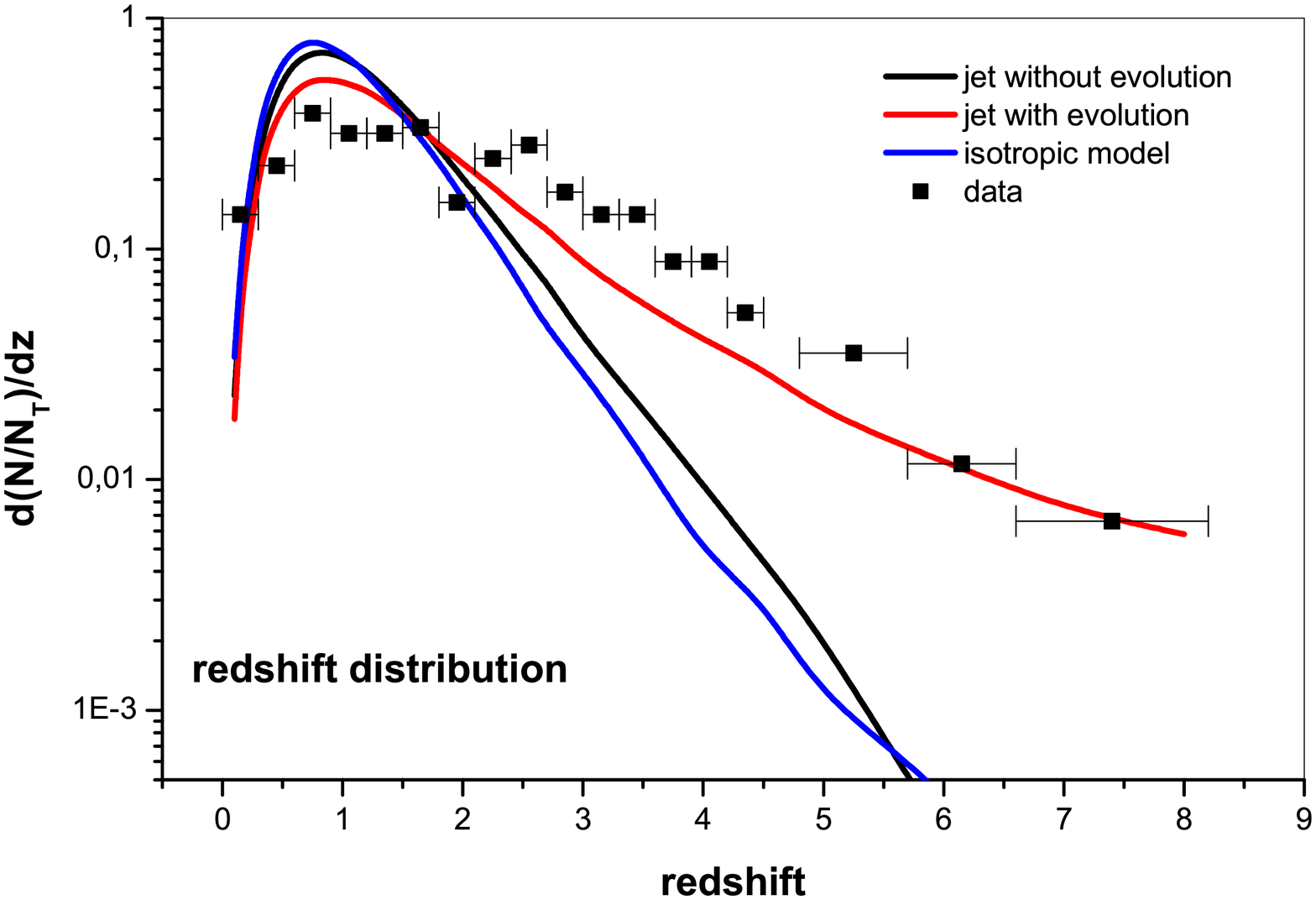}
\caption{Frequency distribution of LGRB redshifts. Data are from online Swift-BAT integrated parameter and 
error bars indicate the bin width.
The different curves correspond to distributions derived from the present simulations: isotropic (blue curve) and
jet (black curve) models without evolution and jet model including evolution (red curve).}
\end{figure}
In figure 7 we show the simulated redshift distributions for the three cases considered in the present investigation
and data for 189 events. As expected, the isotropic and the jet models without evolution underestimate the
frequency of events for $z >$2. Including a moderate evolution in the jet energy distribution permits
a better representation of data, in agreement with previous investigations. Note that this evolutionary jet model
predicts a fraction of events for $z\geq 6$ equal to 0.022, which corresponds to four events for a sample of 189 objects,
in good agreement with the fact that four events with $z \geq 6$ are present in the considered sample.

\subsection{Supernova connection}

Observations seem to suggest that at least a fraction of LGRBs are linked to type Ibc supernova, whose 
spectra are dominated by broad absorption lines. Most of the associated bursts have isotropic energies of
about $10^{49}$ erg and are assumed to have jets with wide opening angles ($E_J \sim E_{iso}$).
It is not clear yet if these low luminosity bursts (LL GRBs) constitute a particular subclass of the LGRBs
\citep{2007ApJ...662.1111L, 2007ApJ...657L..73G}. According to estimates by \citet{2006Natur.442.1014S}, the
rate of LL GRBs is about 230 $Gpc^{-3}yr^{-1}$, similar to the rate currently derived for LGRBs. However,
\citet{2007ApJ...657L..73G} concluded that  energetic LGRBs represent only a small fraction of LL GRBs. This
clearly is an open question.

The fraction by mass $\lambda_{grb}$ of formed stars giving origin to LGRBs was estimated in the present study
to be $\lambda_{grb}=2.8\times 10^{-5}~M_{\odot}^{-1}$. If the initial mass function $\zeta(m)$ is known, $\lambda_{grb}$
fixes the lower mass limit $m_{grb}$ of progenitors of LGRBs by the relation
\begin{equation}
\label{imf}
\lambda_{grb}=\int^{m_u}_{m_{grb}}\zeta(m)dm~.
\end{equation}
Using the initial mass function by \citet{1979ApJS...41..513M} normalized in the mass interval $0.1-125~M_{\odot}$, from the
equation above, one obtains that the progenitors of LGRBs must have a minimum mass of about 90 $M_{\odot}$. On the other hand,
\citet{2009A&A...502..611G} estimated based on rotating stellar models computed by \citet{2003A&A...404..975M, 2005A&A...429..581M} that for solar 
metallicities the minimum mass of the progenitors of type Ic supernovae is about 39 $M_{\odot}$. Thus, from eq.~\ref{imf} one
obtains $\lambda_{Ic}\simeq 3.1\times 10^{-4}~M_{\odot}^{-1}$
for the fraction by mass of formed stars that become type Ic supernovae, .
 These values imply that only $\sim$ 9.0\% of SNIc could be
associated with LGRBs, namely, only those with a very massive progenitor. Instead of using lower mass limits derived
theoretically, it is possible to directly compare the present derived formation rate of LGRBs with the SNIbc rate estimated
from observations. Using data by \citet{1999A&A...351..459C} for rates of core-collapsed supernovae and the ratio between different
supernova types derived by \citet{2008ApJ...673..999P}, we estimated that $R_{SNIbc}\sim 2.7\times 10^4~Gpc^{-3}yr^{-1}$. Comparing
this with the LGRB rate derived in this work, one concludes that only $\sim$ 1\% of type Ic supernovae could be related with LGRBs.
Note that the SNIbc rate estimated from observations implies progenitor masses lower than those derived theoretically, that is
around 16 $M_{\odot}$. However, these results indicate that the majority of SNIbc are not related to 
LGRBs, in agreement with
the conclusions by \citet{2006ApJ...638..930S}. However, it is worth mentioning that recent radio observations 
of two SNIbc supernovae \citep{2010Natur.463..513S}
and 2007gr \citep{2010Natur.463..516P} not associated with a $\gamma$-flash indicate the presence of mildly relativistic 
outflows, which require a central engine able to power 
the observed flows.

We emphasize that if the progenitors of LGRBs are not single stars, our analysis
is invalid. This is also true if the initial mass function (IMF) evolves as assumed by \citet{2011ApJ...727L..34W}, who
proposed an IMF that becomes increasingly top-heavy at high reshift, producing higher LGRB rates at high $z$.

\section{Conclusions}

We reported results based on Monte Carlo simulations aimed at studying the properties of LGRBs. Mock catalogs of 
LGRBs derived from these simulations permitted us to predict observed fluences, and from the comparison
with actual data, we were able to estimate the parameters that define the energy distribution of LGRBs. These
simulations permitted us also to compute the visibility function that measures the probability for a given burst
generated at a given redshift $z$ to be able to trigger the Swift-BAT detector. 

When an isotropic emission model without evolution is considered, the observed fluence distribution derived from the
Swift-BAT catalog can be reproduced by an energy distribution modeled by a log-normal characterized by a
median $\log E_{iso}$=50.95 (erg) and a dispersion $\sigma_{\log E}$=0.92. When a comparison is made
with the mean energy derived directly from observations, we realize that due to the Malmquist bias, the
aforementioned intrinsic (simulated) value is about 60 times lower. The local formation rate of LGRBs
derived from our simulations under the assumption of isotropic emission is $R_{grb}$=1.6 $Gpc^{-3}yr^{-1}$,
an intermediate value when compared with those derived by \citet{2007ApJ...656L..49S} and \citet{2008A&A...491..157P}. 
The present local formation rate based on the isotropic emission model agrees with the result 
by \citet{2010MNRAS.406.1944W}, but we emphasize  that in their approach evolution effects are present in 
the luminosity function of GRBs. The formation rate of subluminous
LGRBs (under the assumption of isotropic emission) could be considerably higher. \citet{2013MNRAS.428..167H}
 derived a formation rate of $\sim$ 150 $Gpc^{-3}yr^{-1}$ for these events. 

For beamed emission model without evolution of the energy distribution is considered, the observed fluence
distribution can be fitted by simulated data when a log-normal defined by a median $\log E_J$=49.26 (in erg) and a dispersion
$\sigma_{\log E}$=0.43 represent the intrinsic energy distribution of LGRBs. As expected, these results considerably alleviate
the energy requirements for GRBs models. The resulting local formation rate for this model is $R_{rgb}$=324
$Gpc^{-3}yr^{-1}$. \citet{2001ApJ...562L..55F} obtained a local formation rate in the context of the jet model of 
$R_{grb} \sim$ 250 $Gpc^{-3}yr^{-1}$, similar to our result this agreement is fortuitous
for the following reason: they took the isotropic rate given by \citet{2001ApJ...552...36S} and corrected this value by the
harmonic mean of the beaming fraction. A similar reasoning was given by \citet{2005ApJ...620..355L} in their estimates
of the local formation rate. 

The redshift distribution predicted by the isotropic and the jet models without evolutionary effects underestimates
the number of LGRBs occurring at $z >$ 2, in agreement with previous investigations concluding 
the necessity of a moderate/strong evolution
of the luminosity function. The model prediction of the observed
redshift distribution (currently including 189 objects) can be considerably improved
if the jet energy distribution evolves in the sense that at high redshift bursts are more energetic. Such an
evolution was modeled in our simulations by assuming that the median of the log-normal describing the energy
distribution varies as $E_{med,J}\propto e^{\delta(1+z)}$. Different set of simulations were performed by varying the slope 
$d\lg E_{med}/d(1+z)$=$\delta$, and the best representation was found for $\delta$=0.5 (see figure 7). The 
ratio between the mean jet energies for events produced 
at $z=7$ and $z=1$ is about 20, indicating a moderate evolution in the energy distribution. For this model,
the derived local formation is $R_{grb}$=290 $Gpc^{-3}yr^{-1}$ and the expected average jet energy is
$3.0\times 10^{49}$ erg, in agreement with the value derived by \citet{2012ApJ...745..168L}. Nevertheless, these
authors have considered the opening angle to be related to the redshift, while in our simulations both
quantities were statistically independent.

The aforementioned evolutionary jet model led to a fraction by mass of formed stars originating in LGRBs of
$\lambda_{grb}=2.8\times 10^{-5}~M_{\odot}^{-1}$. If the IMF does not evolve and a \citet{1979ApJS...41..513M}
IMF is adopted, the derived value of $\lambda_{grb}$ implies that the minimum mass to produce an LGRB is
$90~M_{\odot}$, indicating that only very massive stars are associated to these events. Moreover,
the ratio between LGRBs and the SNIbc formation rates is in the range 
$0.01-0.09$, supporting the idea that stars less massive than the above limit may produce an SNIbc event, but not 
necessarily an LGRB.

\begin{acknowledgements}
CK acknowledges the financial support from the University of Nice-Sophia Antipolis for her PhD project.
The authors also thank J.-L. Atteia and T. Regimbau for the critical reading of the manuscript.
\end{acknowledgements}


\bibliographystyle{aa} 
\bibliography{Cref} 

\end{document}